\documentclass[aps,prb]{revtex4}
\input{epsf}
\begin{document}
\title{Properties of 1D two-barrier quantum pump with harmonically
oscillating barriers}

\author{M.M. Mahmoodian}
\email{mahmood@isp.nsc.ru}
\affiliation{Institute of Semiconductor
Physics, Siberian Division, Russian Academy of Sciences,
\\Novosibirsk, 630090 Russia}

\author{L.S. Braginskii}
\email{brag@isp.nsc.ru}
\affiliation{Institute of Semiconductor
Physics, Siberian Division, Russian Academy of Sciences,
\\Novosibirsk, 630090 Russia}

\author{M.V. Entin}
\email{entin@isp.nsc.ru}
\affiliation{Institute of Semiconductor
Physics, Siberian Division, Russian Academy of Sciences,
\\Novosibirsk, 630090 Russia}


\begin{abstract}
We study a one-dimensional quantum pump composed of two
oscillating delta-functional barriers. The linear and non-linear
regimes are considered. The harmonic signal applied to any or both
barriers causes the stationary current. The direction and value of
the current depend on the frequency, distance between barriers,
value of stationary and oscillating parts of barrier potential and
the phase shift between alternating voltages.
\end{abstract}

 \maketitle

The quantum pump or a device which generate a stationary current
under action of alternating voltage was a subject of numerous
recent publications (for example,
\cite{Moskalets03}-\cite{Lotkhov}). The quantum pump is
essentially analogous to various versions of the photovoltaic
effect, studied in detail from the beginning of the 1980s
\cite{Bel}-\cite{Iv}. The difference is that the photovoltaic
effect is related to the emergence of a direct current in a
homogeneous macroscopic medium (the only exception is the
mesoscopic photovoltaic effect), while a pump is a microscopic
object. From the phenomenological point of view, the emergence of
a direct current in the pump is not surprising since any
asymmetric microcontact can rectify ac voltage. However, analysis
of adiabatic transport in a quantum-mechanical object leads to new
phenomena, such as quantization of charge transport
\cite{Thouless}. The  quantum pump is a sample of phenomena
important in living matter such as active ion transport through
the cell membrane and bacterial motion (biological motors).

In the recent paper \cite{recent} we have studied the
one-dimensional quantum pump with two oscillating delta-like
potential barriers or wells. We have found a variety of regimes of
the pump operation, depending on the system parameters. In this
paper we continue this study, concentrating on the non-considered
cases, aimed especially at low-frequency and nonlinear operation
modes of the electronic pump.

\subsection*{Basic Equations}
\begin{figure}[ht] \centerline{\epsfysize=7cm
\epsfbox{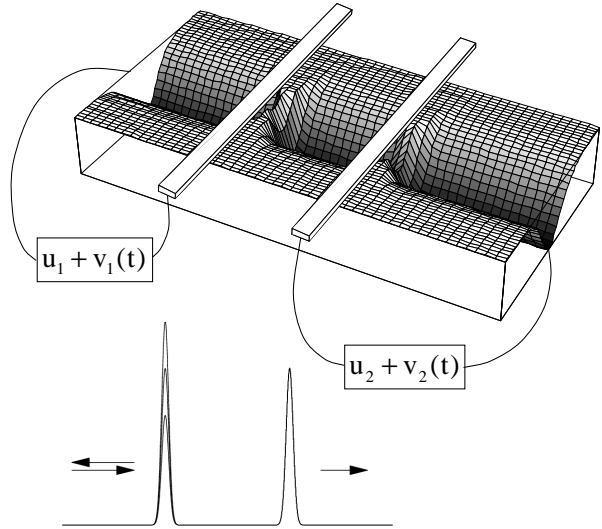}} \caption{Up: The draft of potential of the
pump. Down: One-dimensional model of the pump.}\label{device}
\end{figure}

We study a one-dimensional system with a potential (Fig.
\ref{device})
\begin{equation}\label{1}
U(x)=(u_1+v_1(t))\delta(x+d)+(u_2+v_2(t))\delta(x-d),
\end{equation} where $v_1(t)=v_1 \sin(\omega t),~~~v_2(t)=v_2 \sin(\omega
t+\varphi)$, $2d$ is the distance between $\delta$-shaped barriers
(wells); quantities $u$ and $v$ are measured in units of
$\hbar/md$ ($m$ is the electron mass); momentum $p$ is measured in
units of $\hbar/d$; energy $E$ is measured in units of
$\hbar^2/2md^2$; and frequency is measured in units of
$\hbar/2md^2$. In the absence of an ac signal, the system has two
barriers for positive values of $u_1$ and $u_2$ and two wells for
negative values of these parameters. This system may be considered
as a quantum wire with two narrow gates (see Figure \ref{device})
to which alternating voltages are applied. A direct current can be
induced only in an asymmetric system. The specific direction in
this system is determined by any of factors: the difference of
static voltages $u_1$ and $u_2$, alternating voltages $v_1$ and
$v_2$ or the phase shift between alternating voltages. Unlike
diode system, the alternating voltages are applied to the pump by
the capacitive method.

We assume that the electron gas is in equilibrium and the
distribution functions are identical in the regions $x<-d$ and
$x>d$. The problem is to determine the direct current induced by
the ac field.

The solution to the Schr\"{o}dinger equation with the potential
(\ref{1}) is searched in the form
\begin{eqnarray}\label{wf}
\psi=\sum\limits_n e^{-i(E+n\omega)t}\times\left\{
\begin{array}{rl}\delta_{n,0}e^{ip_nx/d}+r_ne^{-ip_nx/d} & x<-d, \\
a_ne^{ip_nx/d}+b_ne^{-ip_nx/d} & -d<x<d, \\
T_ne^{-i(p+p_n)}e^{ip_nx/d} & x>d.
\end{array}\right.& &
\end{eqnarray}
Here, $p_n=\sqrt{p^2+n\omega}$ and $p=\sqrt{E}$. The wavefunction
(\ref{wf}) corresponds to the wave incident on the barrier from
the left. (In the final formulas, we  mark directions of incident
waves by indices ''$\rightarrow$'' and ''$\leftarrow$'').
Quantities $T_n$ and $r_n$ give the amplitudes of transmission
(reflection) with absorption (for $n>0$) or emission (for $n<0$)
of $n$ ac field quanta, while quantity $T_0$ determines the
amplitude of the elastic process. If the value of $p_n$ becomes
imaginary, the waves moving away from the barriers should be
treated as damped waves, so that $\mbox{Im}p_n>0$.

The transmission amplitudes $T_n$ obey the  equations
\begin{eqnarray}\label{sis}
v_1v_2g_{n-1}e^{-i\varphi}T_{n-2}^\rightarrow-i\left[v_1S_{n-1}+v_2V_ne^{-i\varphi}\right]T_{n-1}^\rightarrow-
\bigg[2W_n+v_1v_2\Big(g_{n-1}e^{i\varphi}+g_{n+1}e^{-i\varphi}\Big)\bigg]T_n^\rightarrow+\nonumber &&\\
i\left[v_1S_{n+1}+v_2V_ne^{i\varphi}\right]T_{n+1}^\rightarrow+v_1v_2g_{n+1}e^{i\varphi}T_{n+2}^\rightarrow=
2ip\delta_{n,0},&& \nonumber\\
v_1v_2g_{n-1}e^{-i\varphi}T_{n-2}^\leftarrow-i\left[v_1S_n+v_2V_{n-1}e^{-i\varphi}\right]T_{n-1}^\leftarrow-
\bigg[2W_n+v_1v_2\Big(g_{n-1}e^{-i\varphi}+g_{n+1}e^{i\varphi}\Big)\bigg]T_n^\leftarrow+
&&\nonumber\\
i\left[v_1S_n+v_2V_{n+1}e^{i\varphi}\right]T_{n+1}^\leftarrow+v_1v_2g_{n+1}e^{i\varphi}T_{n+2}^\leftarrow=
2ip\delta_{n,0}. &&
\end{eqnarray}
Here, $g_n=\sin2p_n/p_n$,
\begin{eqnarray}\label{func}
&&S_n=2u_2g_n+e^{-2ip_n},~V_n=2u_1g_n+e^{-2ip_n},\\
&&W_n=2u_1u_2g_n+(u_1+u_2-ip_n)e^{-2ip_n}.
\end{eqnarray}
Provided that the electrons from the right and left of the contact
are in equilibrium, and they have identical chemical potentials
$\mu$, the stationary current is
\begin{equation}\label{cur}
J=\frac{e}{\pi\hbar}\int
dE\sum_n(|T_n^\rightarrow|^2-|T_n^\leftarrow|^2)f(E)\theta(E+n\omega),
\end{equation}
where $f(E)$ is the Fermi distribution function, and $\theta(x)$
is the Heaviside step function. The current is determined by the
transmission coefficients with real $p_n$ only.

At a low temperature, it is convenient to differentiate the
current with respect to the chemical potential $\mu$:
\begin{equation}\label{G}
{\cal G}=e\frac{\partial}{\partial \mu}J=G_0
\sum_n\theta(\mu+n\omega)(|T_n^\rightarrow|^2-|T_n^\leftarrow|^2)_{p=p_F}.
\end{equation}
Here, $G_0=2e^2/h$ is the conductance quantum, $h$ is the Planck
constant, and $p_F$ is the Fermi momentum. The resultant quantity
${\cal G}$ can be treated as a two-terminal photoconductance (the
conductance for simultaneous change of chemical potentials of
source and drain).

\subsection*{The asymptotic cases}

Let us consider the limit $v_1,v_2\ll u_1,u_2$. The steady-state
problem gives the transmission amplitude
\begin{eqnarray}\label{stat}
T_0=-\frac{ip}{W_0}=-\frac{ip^2}{2u_1u_2\sin2p+(u_1+u_2-ip)pe^{-2ip}},
~~~~~T_n|_{n\neq 0}=0.
\end{eqnarray}
The scattering amplitude vanishes for $p\to 0$ and experiences
oscillations with a period $\delta p=\pi/2$. For large values of
$u_{l,2}$, quantity $T_0$ has poles in the vicinity of points
$p=\pi n/2$.

In the zeroth order of perturbation theory, the direct and reverse
transmission coefficients coincide; consequently, the current
vanishes. The current appears only in the second order of
perturbation theory. Second-order corrections to the current come
only from quantities $T_0$, $T_1$, and $T_{-1}$. Expanding in the
ac signal, we obtain
\begin{eqnarray}\label{j}
&&{\cal G}=G_0\frac{p^2}{4|W_0|^2}
\Bigg\{v_1^2\left(\frac{|S_0|^2-|S_{-1}|^2}{|W_{-1}|^2}\theta(\mu-\omega)
+\frac{|S_0|^2-|S_1|^2}{|W_1|^2}\right)-\nonumber\\
&&v_2^2\left(\frac{|V_0|^2-|V_{-1}|^2}{|W_{-1}|^2}\theta(\mu-\omega)+
\frac{|V_0|^2-|V_1|^2}{|W_1|^2}\right)+\\
&&2v_1v_2~\mbox{Re}\left[\frac{S_0V_{-1}^*-S_{-1}V_0^*}{|W_{-1}|^2}e^{-i\varphi}\theta(\mu-\omega)
+\frac{S_0V_1^*-S_1V_0^*}{{|W_1|^2}}e^{i\varphi}\right]+\nonumber\\
&&4v_1v_2\sin\varphi~\mbox{Im}\left[\frac{S_0V_0-S_{-1}V_{-1}}{W_0W_{-1}}-
\frac{S_0V_0-S_1V_1}{W_0W_1}+2\frac{g_{-1}-g_1}{W_0}\right]\Bigg\}_{p=p_F}.\nonumber
\end{eqnarray}
In the particular case when $u_1=u_2$, the functions $S_n$ and
$V_n$ coincide, and the expression (\ref{j}) obtains the form
\begin{eqnarray}\label{j1}
&&{\cal G}=G_0\frac{p^2}{4|W_0|^2}
\Bigg\{\left(v_1^2-v_2^2\right)\left(\frac{|S_0|^2-|S_{-1}|^2}{|W_{-1}|^2}\theta(\mu-\omega)+
\frac{|S_0|^2-|S_1|^2}{|W_1|^2}\right)+\nonumber\\
&&4v_1v_2\sin\varphi~\mbox{Im}\left[\frac{S_0S_{-1}^*}{|W_{-1}|^2}\theta(\mu-\omega)
-\frac{S_0S_1^*}{{|W_1|^2}}+\frac{S_0^2-1}{W_0}\left(\frac{1}{W_{-1}}-
\frac{1}{W_1}\right)\right]\Bigg\}_{p=p_F}.
\end{eqnarray}
The current is determined by the corrections $T_{\pm1}$ associated
with real emission (absorption) of a single photon. In addition, a
correction to $T_0$ associated with the effect of a virtual
single-photon process on the nonradiative channel also exists.
Apart from the squares of ac signals $v_1$ and $v_2$, the result
for the regime $u_1=u_2$ contains a bilinear combination;
consequently, it is insufficient to consider the response only at
one of the signals. The latter contribution is sensitive to the
relative phase of the signals.

In the case of the large $u_1$ and $u_2$ compared with the Fermi
momentum, the expression (\ref{j}) yields
\begin{eqnarray}\label{j1-1}
&&{\cal
G}=G_0\frac{p^2v_1^2\sin\varphi}{8u_1^7g_0}\left\{\frac{\left(
3p_{-1}-p\right)\theta(\mu-\omega)}{g_{-1}}-
\frac{3p_1-p}{g_1}\right\}_{p=p_F}\stackrel{\small\omega\to 0}
{\longrightarrow}G_0\frac{\omega
p_F^3v_1^2\sin\varphi\left[4p_F\cot
2p_F-5\right]}{8u_1^7\sin^22p_F}.
\end{eqnarray}

If $u_1=u_2=0$,
\begin{eqnarray}\label{j2}
&&{\cal G}=G_0v_1v_2 \sin\varphi\left\{\frac{\sin
2(p-p_{-1})}{p_{-1}^2}\theta(\mu-\omega)+\frac{\sin 2(p_{1}-p)}{p_{1}^2}\right.\nonumber\\
&&\left.+\frac{2\sin2p}{p}\left(\frac{\cos
2p_{-1}}{p_{-1}}\theta(\mu-\omega)-\frac{\cos
2p_1}{p_1}\right)\right\}_{p=p_F}.
\end{eqnarray}
The expression (\ref{j2}) tends to infinity at the single photon
emission threshold. This singularity can be explained by the
resonance with the state of an electron with zero energy: such an
"immobile" state can be interpreted as a bound state.

In addition to the above-mentioned  oscillations with period
$\delta p=\pi/2$, the transmission amplitude experiences
oscillations with periods $\delta p_{\pm 1}=\pi/2$. It can be seen
from expression (\ref{j}) that the extrema in the dependence of
the current on $p$ are located in the vicinity of the points
corresponding to the minima of functions $W_0$ and $W_{\pm1}$ and
are connected with the elastic process as well as with the process
involving the absorption or emission of a field quantum. For
$v_2=0$($v_1=0$), the expression for the current contains only one
term proportional to $v_1^2$($v_2^2$).

For $u_1,u_2\gg p$ the oscillations are transformed into sharp
peaks corresponding to the transmission resonances. For $p\sim 1$,
the transmission amplitude has a characteristic scale of $p\sim
u_1,u_2$. The corresponding structure for small values of $u_1$
and $u_2$ can be treated as a resonance at zero energy. For
negative values of $u_1$ and $u_2$, resonance at bound states
exist (at one or two such states depending on the distance between
the wells).

\subsection*{Numerical results.}

\begin{figure}[ht]
\centerline{\epsfysize=7cm \epsfbox{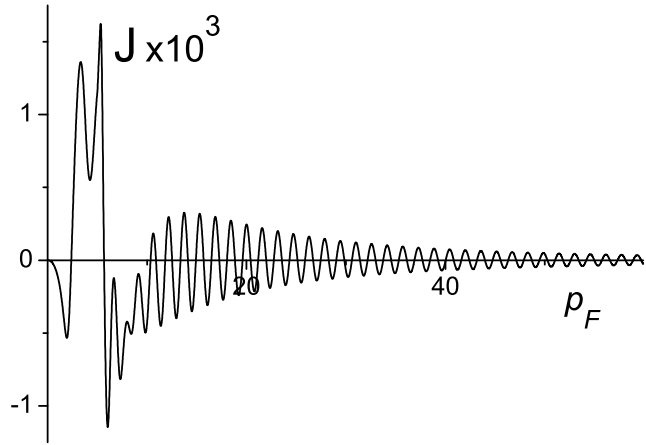}} \caption{The
dependence of the stationary current $J$ (in units $e\hbar/2\pi
md^2$) on the Fermi momentum in a symmetric structure
$u_1=u_2=-1,~v_1=v_2=0.1,~\omega=25,~\varphi=\pi/2$.}\label{J-fermi}
\end{figure}

The Figure \ref{J-fermi} shows the dependence of the stationary
current $J$ on the Fermi momentum in a symmetric structure with
two $\delta$-wells ($|u_1|=|u_2|\gg v_1=v_2,~\varphi=\pi/2$). The
current oscillate with the Fermi momentum with the period $\pi/2$.
These oscillations are related to the resonance at
quasi-stationary states between the wells. The threshold
singularity at $p_F=5$ is associated with zero-energy one-photon
resonance.

\begin{figure}[ht]
\centerline{\epsfysize=7cm \epsfbox{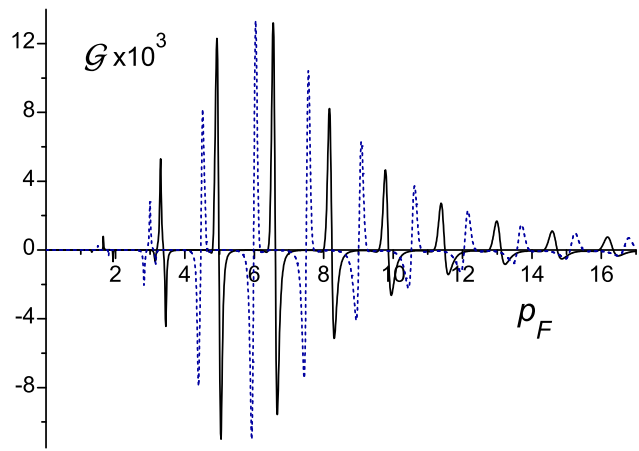}} \caption{The
dependence of $\cal G$ on the Fermi momentum in a symmetric
structure $u_1=u_2=\pm 1,~v_1=v_2=0.1,~\omega=1,~\varphi=\pi/2$.
The solid and dashed curves corresponds to $u_1=1$ and $u_1=-1$,
accordingly.}\label{G-fermi}
\end{figure}

The Figure \ref{G-fermi} demonstrates the dependence of the
quantity $\cal G$ on the Fermi momentum in the symmetric structure
with two identical $\delta$-wells and - $\delta$-barriers. These
cases differs by the sign of the quantity $\cal G$ and by the
small relative shift of the position of the resonance
singularities. Really, within the limits of large $u_1=u_2$ at
$\omega\to 0$ the quantity ${\cal G}\propto u_1^{-7}$
(\ref{j1-1}), i.e. is odd function of amplitude $u_1$ and
accordingly, changes the sign with the changing of the $u_1$ sign.
The  shift of the position of the resonance singularities is
connected with the difference of quasi-stationary energy levels
in these cases.

\begin{figure}[ht]
\centerline{\epsfysize=7cm \epsfbox{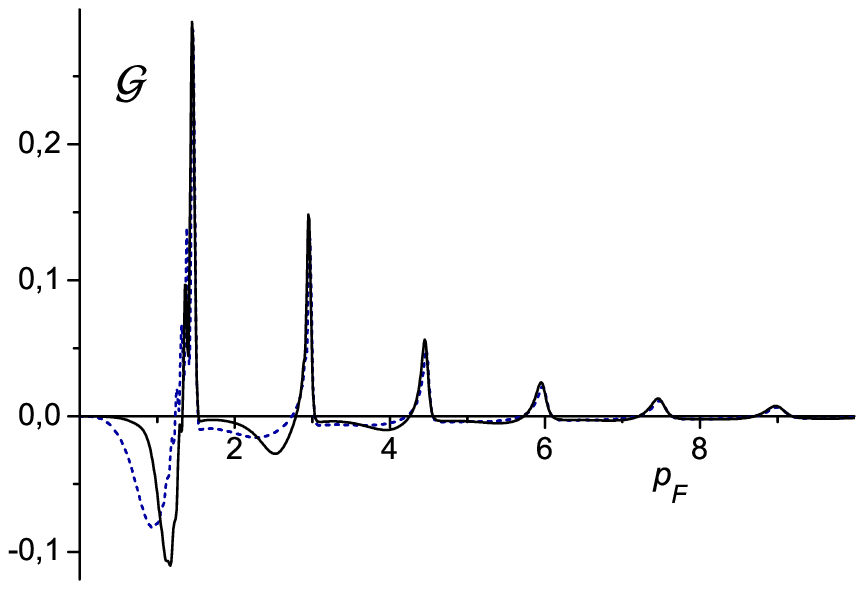}} \caption{The
dependence of $\cal G$ on the Fermi momentum
$u_1=u_2=v_1=v_2=5,~\omega=0.1$ for $\varphi=\pi/2$ (solid curve)
and $\varphi=\pi/3$ (dashed curve).}\label{G-phase}
\end{figure}

The Figure \ref{G-phase} depicts $\cal G$ as a function of Fermi
momentum for two values of the phase $\varphi$ in the symmetric
device. It demonstrates that $\cal G$ is phase-sensitive for small
$p_F$ up to $p_F\sim 5$.  The change of phase modifies the curve,
in particular visibly shifts the first deep. For large $p_F>5$ the
curves correspond to the perturbative expression (\ref{j1}).
\begin{figure}[ht]
\centerline{\epsfysize=7cm \epsfbox{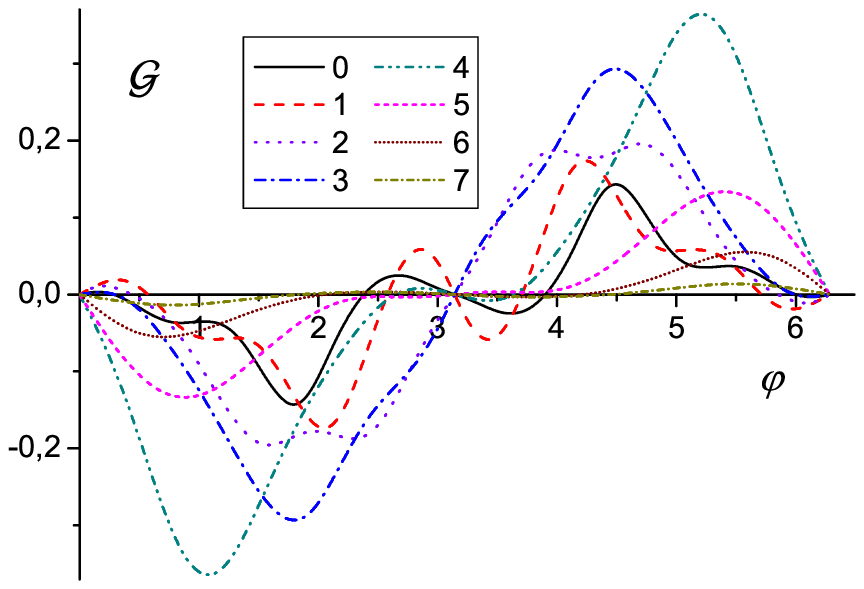}} \caption{The
dependence of $\cal G$ on the phase shift $\varphi$ in a symmetric
structure $\omega =1,~p_F=2,~v_1=v_2=5,~u_1=u_2=0, 1, 2, 3, 4, 5,
6, 7.$}\label{G-phase1}
\end{figure}

\begin{figure}[ht]
\centerline{\epsfysize=7cm \epsfbox{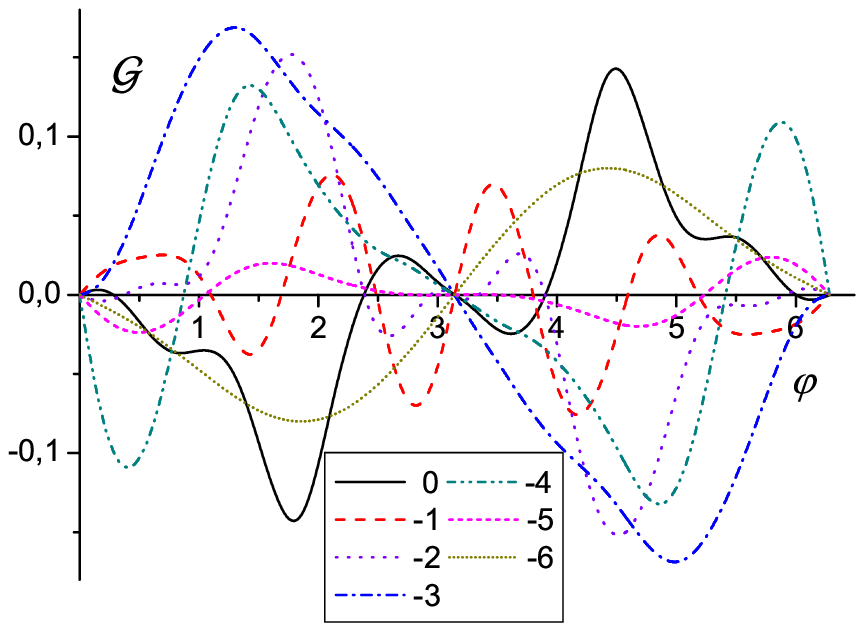}} \caption{The
dependence of $\cal G$ on the phase shift $\varphi$ in a symmetric
structure $\omega =1,~p_F=2,~v_1=v_2=5,~u_1=u_2=0, -1, -2, -3, -4,
-5, -6.$}\label{G-phase2}
\end{figure}
The Figures \ref{G-phase1} and \ref{G-phase2} show the evolution
of the function $\cal G(\varphi)$  in the symmetric device with
two barriers (\ref{G-phase1}) or two wells (\ref{G-phase2}) with
the value of alternating signal at a fixed $p_F=2$. The case of
large $u_{1,2}\gg v_{1,2}$ corresponds to the perturbative
expression (\ref{j1}). This explains the approximative sinusoidal
dependence of $\cal{G}$ on the phase for $|u_{1,2}|>5$. For
relatively small Fermi momenta $p_F<v_{1,2}$, and if also
$u_{1,2}\leq v_{1,2}$, the harmonic (sinus-like) dependence of
$\cal G(\varphi)$ is superimposed on the short-period ($\pi/2$)
oscillations conditioned by the resonance in 4th order of the
perturbation theory.

\begin{figure}[ht]
\centerline{\epsfysize=7cm \epsfbox{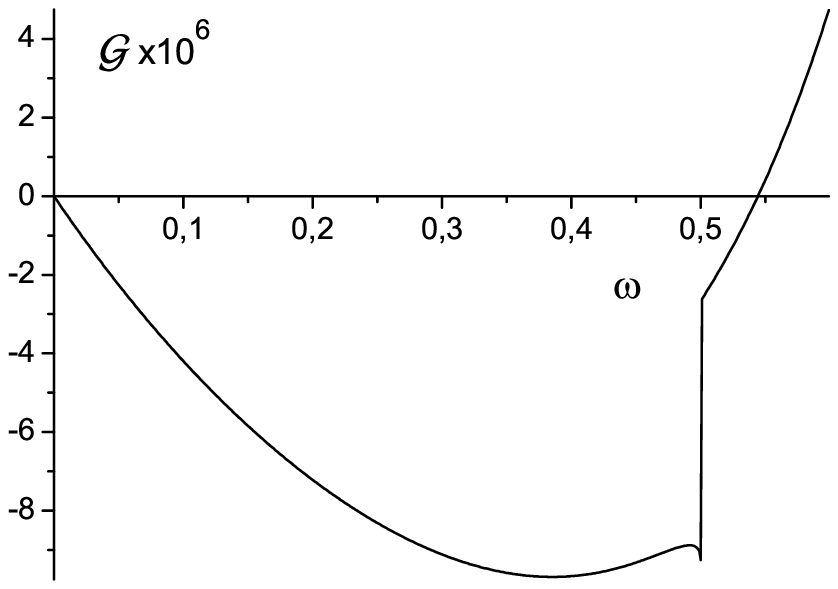}} \caption{The
dependence of $\cal G$ on the frequency
$u_1=1,~u_2=3,~v_1=v_2=0.1,~\varphi=\pi/2,~p_F=0.71$.}\label{G-freq}
\end{figure}

The Figure \ref{G-freq} demonstrates the dependence of $\cal G$ on
the frequency of the alternating signal in the low-frequency
limit. The linear dependence of $\cal G$ in the low-frequency
limit agrees with (\ref{j1-1}). The threshold singularity at
$\omega=0.5$ is related to zero-energy one-photon resonance.

\begin{figure}[ht]
\centerline{\epsfysize=7cm \epsfbox{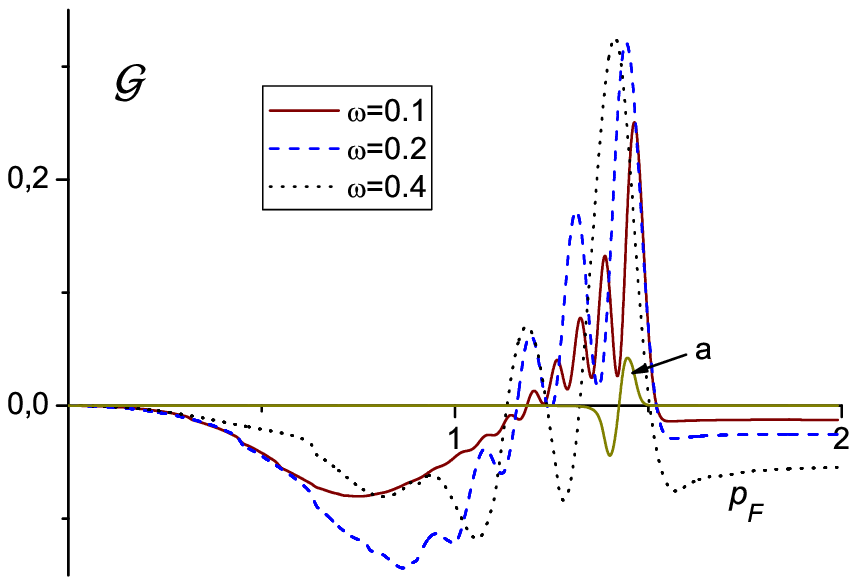}}
\caption{The dependence of $\cal G$ on the Fermi momentum for
different small frequencies (shown in the figure);
$u_1=u_2=v_1=v_2=5,~\varphi=\pi/4$. The curve a) represents low-
signal result for
$u_1=u_2=5,~~v_1=v_2=1,~\varphi=\pi/4,~\omega=0.1$.}\label{G-Fermi-freq}
\end{figure}

The Figure \ref{G-Fermi-freq} depicts $\cal{G}$ for strong
low-frequency
 alternating voltages. The resonance at $p_F=\pi/2$, which
presents in the low-signal regime (see the curve a) obtains the
photon repetitions. They overlap composing damped (with the number
of photons) oscillations. The oscillations rarefy with the
increase of the frequency.

\section*{Conclusions}
The problem of stationary current induced by harmonic signals
applied via two gates to one-dimensional system was studied. The
considered system is described by the simplified double
delta-functional time-dependent barriers. The regimes of weak and
strong external voltage were considered. The current experiences
oscillations as a function of chemical potential. These
oscillations turn into interference resonances if the stationary
barriers or the alternating voltages are strong enough. The
resonances have many-photon nature. The current depends on the
phase shift between gates.

The work was supported by grants of RFBR Nos. 05-02-16939 and
04-02-16398, Program for support of scientific schools of the
Russian Federation No. 593.2003.2 and the Dynasty Foundation.

\end{document}